\documentclass[aps,prl,reprint,a4paper,superscriptaddress,noeprint,nobibnotes]{revtex4-2}
\usepackage[pdfusetitle,colorlinks,linkcolor=blue,citecolor=blue,urlcolor=blue]{hyperref}
\usepackage[numbered,open]{bookmark}
\usepackage[all]{hypcap}
\usepackage{indentfirst,xparse,xspace,tabularx}
\usepackage{amsmath,amsfonts,amssymb,bm}
\usepackage{xcolor,lineno,booktabs}
\usepackage{graphicx}
\ifxetex\usepackage{fontsetup}\else\usepackage[T1]{fontenc}\fi
\usepackage[UKenglish]{babel}
\usepackage[normalem]{ulem}

\hypersetup{pdfauthor=Jing Wang and Xuchen Lin}

\usepackage[angle-symbol-over-decimal,unit-mode=text,
  table-alignment-mode=format,table-number-alignment=center,
  uncertainty-mode=separate,print-unity-mantissa=false,
  range-units=single,range-phrase=--,
  list-final-separator={, and },list-units=single]{siunitx}

\makeatletter
\usepackage{etoolbox,lineno}
\ifxetex%
  \usepackage{xeCJK}
\else
  \usepackage{CJK}
  \AtBeginDocument{\DeclareSIUnit[quantity-product={}]\degree{\text{\textdegree}}}
\fi
\ProvideDocumentEnvironment{CJK*}{m m}{}{}
\AtBeginDocument{
  \newcommand*{\cjkname}[2]{#1 (#2)}
  \@ifpackageloaded{CJK}{}{
    \@ifpackageloaded{xeCJK}{}{
      \renewcommand*{\cjkname}[2]{#1}
    }
  }
}
\pretocmd{\maketitle}{\csname CJK*\endcsname{UTF8}{gbsn}}{}{}
\apptocmd{\maketitle}{%
  \csname endCJK*\endcsname%
  \bookmarksetup{startatroot}%
}{}{}
\expandafter\renewcommand\csname NAT@sort\endcsname{0}
\patchcmd{\NAT@citex}
{\@citea\NAT@hyper@{%
    \NAT@nmfmt{\NAT@nm}%
    \hyper@natlinkbreak{\NAT@aysep\NAT@spacechar}{\@citeb\@extra@b@citeb}%
    \NAT@date}}
{\@citea\NAT@nmfmt{\NAT@nm}%
  \NAT@aysep\NAT@spacechar\NAT@hyper@{\NAT@date}}{}{}
\patchcmd{\NAT@citex}
{\@citea\NAT@hyper@{%
    \NAT@nmfmt{\NAT@nm}%
    \hyper@natlinkbreak{\NAT@spacechar\NAT@@open\if*#1*\else#1\NAT@spacechar\fi}%
    {\@citeb\@extra@b@citeb}%
    \NAT@date}}
{\@citea\NAT@nmfmt{\NAT@nm}%
  \NAT@spacechar\NAT@@open\if*#1*\else#1\NAT@spacechar\fi\NAT@hyper@{\NAT@date}}
{}{}
\pretocmd{\NAT@citexnum}{\@ifnum{\NAT@ctype>\z@}{\let\NAT@hyper@\relax}{}}{}{}
\NewCommandCopy{\IOpt}{\normalsize}
\renewcommand\normalsize{%
  \@setfontsize\normalsize\@ixpt{11}%
  \abovedisplayskip 8.5\p@ \@plus3\p@ \@minus4\p@
  \abovedisplayshortskip \z@ \@plus2\p@
  \belowdisplayshortskip 4\p@ \@plus2\p@ \@minus2\p@
  \def\@listi{\leftmargin\leftmargini
    \topsep 4\p@ \@plus2\p@ \@minus2\p@
    \parsep 2\p@ \@plus\p@ \@minus\p@
    \itemsep \parsep}%
  \belowdisplayskip \abovedisplayskip
}
\normalsize
\ifx\MakeRobust\@undefined \else
  \MakeRobust\normalsize
\fi
\DeclareRobustCommand\small{%
  \@setfontsize\small\@viiipt{9.5}%
  \abovedisplayskip 6\p@ \@plus2\p@ \@minus4\p@
  \abovedisplayshortskip \z@ \@plus\p@
  \belowdisplayshortskip 3\p@ \@plus\p@ \@minus2\p@
  \def\@listi{\leftmargin\leftmargini
    \topsep 3\p@ \@plus\p@ \@minus\p@
    \parsep 2\p@ \@plus\p@ \@minus\p@
    \itemsep \parsep}%
  \belowdisplayskip \abovedisplayskip
}
\DeclareRobustCommand\footnotesize{%
  \@setfontsize\footnotesize\@viipt\@viiipt%
  \abovedisplayskip 5\p@ \@plus2\p@ \@minus3.5\p@
  \abovedisplayshortskip \z@ \@plus\p@
  \belowdisplayshortskip 2.5\p@ \@plus\p@ \@minus2\p@
  \def\@listi{\leftmargin\leftmargini
    \topsep 2.5\p@ \@plus\p@ \@minus\p@
    \parsep 2\p@ \@plus\p@ \@minus\p@
    \itemsep \parsep}%
  \belowdisplayskip \abovedisplayskip
}
\DeclareRobustCommand\scriptsize{\@setfontsize\scriptsize\@vipt\@viipt}
\setcitestyle{super}
\renewcommand\NAT@citesuper[3]{\ifNAT@swa
  \if*#2*\else#2\NAT@spacechar\fi
  \unskip\kern\p@\textsuperscript{\NAT@@open#1\NAT@@close}%
  \if*#3*\else\NAT@spacechar#3\fi\else #1\fi\endgroup}

\citeautoscript@sw{
  \RenewDocumentCommand{\citet}{om}{%
    \IfNoValueTF{#1}%
    {\ \citeauthor{#2} \citep{#2}}%
    {\ \citeauthor{#2} \citep{#2} (#1)}%
  }
  \RenewDocumentCommand{\Citet}{om}{%
    \IfNoValueTF{#1}%
    {\ \Citeauthor{#2} \citep{#2}}%
    {\ \Citeauthor{#2} \citep{#2} (#1)}%
  }
}{}%
\def\section{%
  \@startsection
  {section}%
  {1}%
  {\parindent}%
  {\z@}%
  {-1em}%
  {\normalfont\normalsize\itshape\maybe@addperiod}%
 }%
\def\subsection{%
  \@startsection
  {subsection}%
  {2}%
  {\parindent}%
  {\z@}%
  {-1em}%
  {\normalfont\normalsize\itshape\maybe@addperiod}%
}%
\newcommand{\maybe@addperiod}[1]{%
  #1\@addpunct{.}%
}

\usepackage{fancyhdr}

\def\headrule{\kern 1mm \hrule width 17cm \kern -1mm}%
\patchcmd{\footnoterule}{%
  .5in}{%
  7cm}
{}{}
\def\fps@figure{htbp}
\def\fps@table{htbp}

\ExplSyntaxOn
\NewDocumentCommand \@strcase { m m }
  {
    \str_case:Vn {#1} {#2}
  }
\ExplSyntaxOff
\patchcmd{\@makecaption}{\hsize}{\dimexpr\hsize-0.7cm\relax}{}{}
\patchcmd{\@makecaption}{%
  \@make@capt@title{#1}{#2}%
}{%
  \@strcase\@captype
    {%
      {figure}{\leftskip=0.35cm\rightskip=0.35cm}
      {table}{\leftskip=0.2cm\rightskip=0.2cm}
    }
  \@make@capt@title{\textbf{#1}}{#2}%
}{}{}
\patchcmd{\@makecaption}{%
  \@make@capt@title{#1}{#2}\par%
}{%
  \@strcase\@captype
    {%
      {figure}{\leftskip=0.35cm\rightskip=0.35cm}%
      {table}{\leftskip=0.2cm\rightskip=0.2cm}
    }
  \@make@capt@title{\textbf{#1}}{#2}\par%
}{}{}
\newcommand{\cplyear}{2024} \newcommand{\cplvol}{41}
 \newcommand{\cplpagenumber}{000000}
 \newcommand{\cplpage}{\cplpagenumber-\thepage}

\def\@hangfrom@section#1#2#3{\@hangfrom{#1#2}{#3}}%
\def\@hangfroms@section#1#2{#1{#2}}%
\pretocmd{\bibliography}{%
  \def\section{\@startsection {section}{1}{-\parindent}%
    {-3.5ex \@plus -1ex \@minus -.2ex}%
    {2.3ex \@plus.2ex}%
    {\normalfont\Large\bfseries}}%
}{}{}
\patchcmd{\frontmatter@titlepage}{%
  \thispagestyle{titlepage}%
}{}{}{}
\patchcmd{\titleblock@produce}{%
  \thispagestyle{titlepage}%
}{}{}{}
\renewcommand*{\frontmatter@title@above}{\vspace*{1mm}}
\apptocmd{\frontmatter@above@affiliation@script}{%
  \addvspace{10\p@}%
}{}{}
\patchcmd{\frontmatter@affiliationfont}{%
  \it}{%
  \sl%
}{}{}
\def\frontmatter@preabstractspace{\baselineskip}
\def\frontmatter@abstractwidth{15.9cm}
\patchcmd{\frontmatter@RRAP@format}{\small}{\IOpt}{}{}
\pretocmd{\frontmatter@RRAP@format}{\vspace*{\baselineskip}}{}{}
\patchcmd{\frontmatter@abstractfont}{\small}{\normalsize}{}{}
\apptocmd{\frontmatter@authorformat}{\IOpt}{}{}
\makeatother

\input{journals.tex}
\input{symb.tex}

\setcitestyle{authoryear,round}
\makeatletter
\renewcommand\section{\@startsection {section}{1}{\z@}{-3.5ex \@plus -1ex \@minus -.2ex}{2.3ex \@plus.2ex}{\normalfont\Large\bfseries}}
\renewcommand\subsection{\@startsection{subsection}{2}{\z@}{-3.25ex\@plus -1ex \@minus -.2ex}{1.5ex \@plus .2ex}{\normalfont\large\bfseries}}
\makeatother

\begin{document}
\title{Recent Developments on the \Hi Gas of Low-Redshift Galaxies\texorpdfstring{\\}{ }Seen by the 21\texorpdfstring{\,}{ }cm Emission Lines}

\author{\cjkname{Jing Wang}{王菁}}%
\email[Corresponding author. Email: ]{jwang_astro@pku.edu.cn}
\affiliation{%
  Kavli Institute for Astronomy and Astrophysics, Peking University, Beijing 100871, China
}

\author{\cjkname{Xuchen Lin}{林旭辰}}%
\affiliation{%
Department of Astronomy, School of Physics, Peking University, Beijing 100871, China
}

\received{1 July 2024}
\accepted{10 September 2024}
\date{10 February 2025}

\begin{abstract}
  As a major interstellar medium, the atomic neutral hydrogen (\Hi) plays an important role in the galaxy evolution.
  It provides the ingredient for star formation, and sensitively traces the internal processes and external perturbations influencing the galaxy.
  With the beginning of many new radio telescopes and surveys, \Hi may make a more significant contribution to the understanding of galaxies in the near future.
  This review discusses the major development of the \qty{21}{\cm} emission-line \Hi observations and studies in the past few years, including its scaling relations with other galaxy properties, its kinematics and structures, its role in environmental studies, and its constraints on hydrodynamical simulations.
  The local-Universe \Hi scaling relations of stellar-mass--selected samples extend smoothly to \qty{1e9}{\Msun} stellar mass, with a tentative evolution to the redshift of \num{\sim0.1}.
  The development of measurement techniques enables better estimations of \Hi non-circular motion, dispersion, and thickness, and new observations revealed extended or extra-planar \Hi structures, both helpfully constraining the gas accretion, stellar feedback, and star formation processes of galaxy evolution models.
  \Hi is very useful for tracing the satellite evolution in dense environments, the studies of which would benefit from ongoing blind \Hi surveys.
  Though simulations still cannot fully reproduce \Hi gas properties, they help to understand the role of possible factors in regulating \Hi properties.
\end{abstract}

\maketitle

\section{Background and Introduction}
Ever since galaxies were recognized as ecosystems regulated or sustained by gas flows, the atomic hydrogen gas (\Hi) has received considerable attention in the field of galaxy evolution.
\Hi is a major interstellar medium (ISM) component, serving as the reservoir of star-forming material for galaxies that are still actively forming stars.
Its relatively cool kinematics makes it a fairly simple tracer of the underlying mass distributions, from which we could infer the dark matter distributions.
Because of its rather tenuous and extended nature compared to other disc components, its distribution and kinematics are sensitive to perturbations from internal feedbacks, disc dynamics, and hydrodynamic and gravitational disturbances from the galaxies beyond.
Its typical temperature of a few hundreds to a few thousands Kelvins suggests its indicative or diagnostic role for the cool component of multi-phase gas filling the vast Universe, particularly in the relatively dense filament and circumgalactic medium (CGM)\@.
These are commonly mentioned reasons that \Hi are potentially useful and important in our understanding of galaxy evolution.
These aspects cover the whole baryonic cycle of gas accretion, star formation, feedback, recycling, and environmental redistribution.
They largely determine the evolutionary fate of galaxies through setting the star-formation capability and quenching potential.

In response to the demand of detecting the \Hi and fully exploiting its diagnostic power, new radio telescopes that emphasize the detection of \qty{21}{\cm} spectral lines have been built globally, with new data in the past decade covering the \Hi Universe with unprecedented resolution, depth, and sky area.
Yet, in this rapidly developing field, the conflict will still be evident and painful for some time, between the need to immediately solve specific scientific problems, and the need to improve the data quality and technological maturity.
A clear view about the frontier and limitations of the field, as well as the major direction where most efforts are being put, should help to spare us from confusion and frustration, and to finally achieve the original intension of developing those observational campaigns.
This review is put together for this purpose for the authors themselves, and hopefully it may also benefit other researchers who are interested in using \Hi data to study galaxy evolution.

This review focuses on the results based on the observation of \qty{21}{\cm} emission line, a convenient tracer of \Hi.
When the atomic hyperfine state with total (electron plus proton) spin~1 decays to the one with total spin~0, a photon is emitted with a wavelength of \qty{21.11}{\cm}.
In most extragalactic studies, we linearly convert \qty{21}{\cm} emission line surface brightnesses and luminosities into \Hi column densities and masses.
However, the emission line observation can be complexed by self-absorption, particularly when the \Hi is optically thick (roughly $\NHi>\qty{1e21}{\per\cm\squared}$) and/or has a bright continuum source along the line of sight (LoS).
Then, the absorption-related parameters, optical depth and spin temperature, need be estimated before deriving the column densities, and one needs two LoSs with and without a background radio continuum source in order to solve for the two variables (see \citealt{2023ARA&A..61...19M} for details).
Sadly, for most external galaxies, the observation capability (sensitivity, spatial resolution, and existence of background radio bright source) limits us from using the absorption line technique, and we have to assume optically thin conditions.
Before sophisticated models are built in the Milky Way (MW) for empirical correction of extragalactic studies, luckily, statistical analysis reveals that the $\NHi<\qty{1e21}{\per\cm\squared}$ \Hi dominates the area coverage/incidence and mass of \Hi \citep{2020ARA&A..58..363P}.
It significantly mitigates the worries of using \qty{21}{\cm} emission line observation for smoothed column densities in most ISM environments, and for most \Hi-integral-mass-based studies.

This review is also an update based on the earlier review by \citet[in Chinese]{2017SSPMA..47d9809W}, which summarized the literature on five topics: the \Hi mass scaling relations, the distribution of \Hi in galaxies, signatures of gas accretion, the relation between \Hi and star formation, and the influence of environmental effects.
Major advances in \Hi science in the past seven years are promoted by the full Arecibo Legacy Fast ALFA Survey (ALFALFA) data release \citep{2018ApJ...861...49H}, the transition of traditional radio instruments towards large programs, the onset of Square Kilometre Array (SKA) pathfinder surveys, including Widefield ASKAP $L$-band Legacy All-sky Blind surveY \citep[WALLABY,][]{2020Ap&SS.365..118K}, Apertif \citep{2022A&A...658A.146V}, and MeerKAT legacy surveys, and the operation of Five-hundred-meter Aperture Spherical radio Telescope \citep[FAST,][]{2019SCPMA..6259502J}.
They also benefited from the advances in techniques and developments of new analysis tools for better source finding \citep{2021MNRAS.506.3962W}, kinematically modelling \citep{2015MNRAS.451.3021D}, and visualization \citep{2017A&C....19...45P}, etc.
Thus, compared to seven years ago, we can look into the integrated \Hi properties with better sample diversities, and the resolved \Hi distribution and kinematics with better statistics.
We are also still far from linking the detailed \Hi distribution and kinematics in an unbiased way to the evolutionary picture established earlier based on optical photometric and spectroscopic surveys, or tracking the depletion of \Hi in galaxies down to \qty{1}{\percent} in \Hi-stellar mass fractions beyond the \qty{50}{\Mpc} Local Universe, or directly tracing the evolution of \Hi (other than $\Omega\sbHi$) back to more than \qty{4}{\Gyr}.
On the other hand, we have seen promising endeavours and encouraging progresses towards these directions.
For the time being, in order to obtain useful constraints from \Hi data, studies from the perspective of simulations or other wavelengths need to carefully correct or control for the selection effects, or to properly trim (down-grade) their own resolution.
There are also continuous efforts on individual systems, exploring physical mechanisms that are poorly understood (e.g.\ turbulence cascade, magnetic field and ISM coupling, gas mixing), digging deeply into the complex interplay of processes driving the galaxy evolution.
These efforts are parallel, and cooperate with the exploration of high-resolution simulations in a similar direction, aiming at formulating empirical constraints or recipes for cosmological simulations.

For a broader context of \Hi science and deeper knowledge on specialized topics, we refer the readers to the many insightful and comprehensive reviews on different aspects of galactic \Hi science in the literature.
In the past ten years, \citet{2018A&ARv..26....4M}, \citet{2021PASA...38...35C}, \citet{2022A&ARv..30....3B}, \citet{2023ARA&A..61...19M}, and \citet{2024arXiv240217004H} have reviewed the \Hi (sometimes together with other ISM components) in active galactic nucleus (AGN)-host galaxies, group galaxies, cluster galaxies, the MW, and dwarf galaxies, respectively.
Based on the stellar mass-selected representative sample of Extended GALEX Arecibo SDSS Survey \citep[xGASS,][]{2018MNRAS.476..875C} and Extended CO Legacy Database for GASS \citep[xCOLD GASS,][]{2017ApJS..233...22S}, \citet{2022ARA&A..60..319S} summarized major lessons learned on star formation and quenching, using scaling relations of integrated \Hi and molecular gas masses.
\citet{2020Ap&SS.365..118K}, \citet{2022A&A...667A..38A}, \citet{2021A&A...646A..35M}, \citet{2023A&A...673A.146S}, and \citet{2024A&A...688A.109D} provided inspiring future perspectives on possible scientific advances with on-going \Hi major campaigns, including WALLABY, Apertif, MeerKAT International GigaHertz
Tiered Extragalactic Exploration survey (MIGHTEE)-\Hi \citep{2021A&A...646A..35M}, MeerKAT Fornax Survey \citep{2023A&A...673A.146S}, and MeerKAT \Hi Observations of Nearby Galactic Objects: Observing Southern Emitters \citep[MHONGOOSE,][]{2024A&A...688A.109D}.
\Citet{2015aska.confE.129D} and \citet{2015aska.confE.167S} shared insight on similar directions by looking further into the SKA era.

For the convenience of presentation, we define low-$z$ Universe where the redshift $z<0.06$, and Local Universe those within \qty{100}{\Mpc}.
The former is roughly the redshift range detected by ALFALFA, and the latter is roughly the distance of the Coma cluster where most interferometry-based resolved studies are conducted.

In the following, we review the recent development in \Hi scaling relations (Section~\ref{sec:scaling}), kinematics (Section~\ref{sec:kinematics}), and structures (Section~\ref{sec:structure}).
In Sections \ref{sec:env} and~\ref{sec:simu}, the \Hi-related environmental studies and the \Hi in contemporary simulations are discussed, respectively.
The summary is given in Section~\ref{sec:summary}.

\section{\Hi Scaling Relations}
\label{sec:scaling}
The \Hi scaling relations statistically link the properties of \Hi to that of other galaxy components, supply clues and reference points for galaxy evolution studies, and provide an easy test for simulations.
Previously, the large-area uniform surveys like ALFALFA made it possible to determine the upper envelope of the \Hi mass-related distributions in the low-$z$ Universe.
Still, because of the fixed \Hi detection limit, these surveys are biased towards \Hi-rich galaxies, and are not stellar mass complete.
The smearing due to the single-dish nature also impedes extending the scaling relations beyond the Local Universe.

\subsection{Global Scaling Relations}
The xGASS \citep{2018MNRAS.476..875C} is the extended version of GASS \citep{2010MNRAS.403..683C}, forming an almost gas fraction-limited sample covering the stellar mass range of $\qty{1e9}{\Msun}<M_*<\qty[parse-numbers=false]{10^{11.5}}{\Msun}$, with the left $M_*$ bound being \qty{1}{\dex} lower.
Their `representative sample' is able to reflect the \Hi property distributions in their $M_*$ and redshift ($0.01<z<0.05$) selection range.

The major scaling relations of the \Hi gas fraction ($\fHi=\MHi/M_*$) found by GASS, including those with $M_*$, stellar surface density $\mu_*:=M_*/(2\pi\Rso^2)$ (where \Rso is the $r$-band half-light radius), specific star-formation rate (sSFR), and the $\NUV-r$ colour, remain and extend smoothly into lower stellar mass \citep{2018MNRAS.476..875C}.
The relation with $\NUV-r$ is still the tightest, which traces the dust-attenuated star formation.
Compared with the one directly with sSFR, the attenuation from dusts that spatially correlate with \Hi reduces the scatter of this relation.
When including the molecular gas mass from xCOLD GASS \citep{2017ApJS..233...22S} and replacing the \fHi with the fraction of total gas, the aforementioned relations become slightly tighter, even though \Hi actually dominates most of the gas mass.
The deviations from these total gas relations strongly correlate with the conversion fraction between molecular and atomic gas, $\Rmol:=\MHZ/\MHi$, which is almost solely driven by the change in \fHi rather than the molecular fraction.
When comparing the logarithmic deviations from their respective main sequence with $M_*$, those of molecular gas are also more tightly related with the deviations \dMS from the star-forming main sequence (SFMS) than the \Hi deviations are \citep{2020MNRAS.493.1982J}, but the general trend is clear such that the decrease of star-formation rate (SFR) correlates with the drop in \Hi content \citep{2020MNRAS.494L..42C}.
The \Hi links to star formation in a less-direct way than the molecular, both because the \Hi needs to firstly convert into molecular gas before stars form near star-forming sites \citep{2008AJ....136.2782L,2018MNRAS.477.2716K,2022ApJ...936..137O}, and because it has a large spatial extension than the star-forming stellar disks at most times \citep{2020ApJ...890...63W}.
On the other hand, in order to sustain the star-forming status, the molecular gas has to be replenished by the \Hi, as its depletion time is \qty[input-comparators=\lesssim]{\lesssim1}{\Gyr} both globally in star-forming galaxies \citep{2017ApJS..233...22S} and locally in star-forming relations\citep{2008AJ....136.2782L};
in the latter case, it can be even shorter if not for the replenishment of \Hi after being destroyed by stellar feedbacks \citep{2022ApJ...936..137O}.
These together suggest the importance of \Hi as star-formation material reservoir, an almost determining role on star-formation sustaining and quenching in galaxies on galaxy evolutionary time-scales \citep[e.g.~\num{0.4} times the Hubble time,][]{2009ApJ...703..785D}, but they also highlight the necessity of understanding the intermediate steps (e.g.~planar radial inflow of \Hi, conversion of \Hi to the molecular) in order for us to physically link \Hi to star formation and galaxy evolution.

\subsection{Stellar-disc-related Inner \Hi}
\begin{figure}
  \centering
  \includegraphics[width=2.5in]{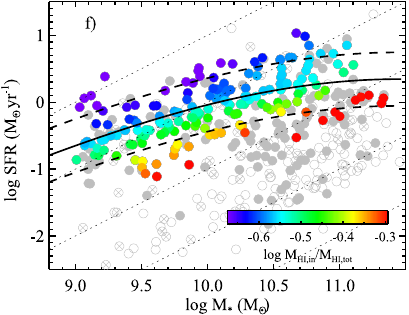}
  \caption{
    The distribution of galaxies on SFR--$M_*$ plane.
    The disc-like galaxies from xGASS with CO detections are colour-coded by the fraction of inner \Hi.
    The inner \Hi fraction anti-correlates with the deviation from the main sequence (MS, solid line), but remains almost constant along the MS\@.
    From the Figure~7 of \citet{2020ApJ...890...63W}.
    \copyright~AAS\@.
  }
  \label{fig:innerHI}
\end{figure}

Based on such an idea, the \Hi disc could be further separated into those within and beyond the optical stellar disc.
Due to a lack of high-resolution \Hi images, \citet{2020ApJ...890...63W} used the total \MHi to deduce the \Hi disc size from the tight \Hi size--mass relation \citep{2016MNRAS.460.2143W}, and divided the \Hi mass at the optical $r$-band \Rqo, the radius enclosing \qty{90}{\percent} of total flux.
When using the inner \Hi instead for \fHi, the relations introduced above obtain a smaller scatter within the same sample, especially the ones with $M_*$ and $\mu_*$.
The \Rmol of inner \Hi exhibits a correlation with \dMS \citep{2020ApJ...890...63W}, which was not found with global \Hi values \citep{2018MNRAS.476..875C}.
Correspondingly (see Figure~\ref{fig:innerHI}), the ratio of inner to total \Hi \emph{anti-correlates} with \dMS, indicating that the efficiency of \Hi inflow assists in reducing the scatter of SFMS, but this efficiency could not be the reason that SFMS flattens at high $M_*$, because this ratio has no dependence on $M_*$ for main-sequence galaxies.
Participating in the later stage of gas-fuelling process, the inner \Hi seems to more closely link to the stellar component and the SF process than the global value does.
Another example is about the gas metallicity, which is diluted by inflowing pristine gas, but could also be enriched by the ensuing star formation.
The more important role of dilution has been statistically confirmed with stacked global \Hi and the fibre metallicity measurements \citep{2018MNRAS.473.1868B}.
\citet{2022ApJ...933...39C} further found that the gas metallicity at the effective radius more strongly anti-correlates with the inner \Hi than both the total \Hi and SFR, directly supporting the scenario that while the gas within the stellar disc is replenished, the metallicity is simultaneously diluted by the accreted gas.
Especially, the relation between the inner \Hi mass and gas metallicity weakens again, if the metallicity is measured at the galaxy central region instead of at the effective radius \citep{2022ApJ...933...39C}.
The importance of being cospatial has also been found for dust mass \citep{2023ApJ...950...84L}, which have better correlations with the inner neural gas (\Hi plus the molecular).
However, no more detailed analyses on the spatial correlation between \Hi, metallicity, and dust could be conducted without enough resolved \Hi images.

\subsection{Relations with Angular Momentum and Morphologies}
With the expansion of \Hi-detected galaxy sample, it is possible to statistically study the relation between the specific angular momentum $j$ (i.e.~normalized by mass) and mass for \Hi and total baryonic mass in disc galaxies.
Different studies \citep{2021MNRAS.507..565K,2021A&A...647A..76M,2021A&A...651L..15M,2021MNRAS.507.2949M,2022MNRAS.509.3751H,2022MNRAS.516.4043H,2024MNRAS.527..931E} reported different slopes between $j$ and $M$ ranging around the value of $2/3$ originally proposed \citep{1983IAUS..100..391F}.
This relation has been explained using a global instability parameter of the baryonic disc that regulates the gas inflow process, and thus the mass fraction of \Hi \citep{2016ApJ...824L..26O}.
In this theory, gases are assumed to be in a quasi-stable state.
The strong correlation between $j$, $M$, and \fHi has been found in larger samples, including xGASS \citep{2022MNRAS.516.4043H} or resolved \Hi observations \citep{2021A&A...651L..15M}, but the exact power-law indices are much different from the prediction of stability theory.
The \fHi of a large portion of galaxies is below the requirement of being quasi-stable \citep{2021MNRAS.507.2949M,2022MNRAS.516.4043H}.
In addition, the $j$--$M$ relation has also been found to have secondary dependence on morphological parameters like bulge-to-total ratio \citep{2022MNRAS.509.3751H} or the environment \citep{2021MNRAS.507.2949M}, indicating possibly different regulators.
\citet{2021A&A...647A..76M} argued that the baryonic $j$--$M$ relation could be explained as a corollary of the one of dark matter halo, and reflects the fraction of angular momentum retained by baryons from the dark matter halo \citep[also see chapter~10 of][]{2020igfe.book.....C}.
Indeed, the much tighter mass--size relation and Tully--Fisher relation could naturally lead to the relation between $j$ and $M$.
We note that the method of measuring angular momentum, especially the radius range, should be carefully taken into account when comparing different studies, as the region with lower luminosity at a rather large radius would still have a large weight on the total specific angular momentum value.

The morphology of the stellar component could also possibly influence the \Hi properties, while the \Hi could alter the stellar structure through regulating star formation and maintaining a cool dynamic status.
A higher $\mu_*$ \citep{2018MNRAS.476..875C} or bulge-to-total ratio \citep{2019MNRAS.490.4060C} is linked with a lower \fHi, against the scenario of morphological quenching \citep{2009ApJ...707..250M}, and when including only the star-forming galaxies, \fHi is independent of the bulge fraction \citep{2019MNRAS.490.4060C}.
The $\mu_*$ could be separated into the contribution from disc and bulge, with the former more tightly correlated with \fHi \citep{2020MNRAS.492.2393C}.
The \fHi has also been found to have little correlation with the bulge colour, but strongly depends on the disc colour \citep{2020MNRAS.492.2393C}.
The strength of stellar spiral arm is reported to correlate with gas mass fraction as well \citep{2021ApJ...917...88Y}.
These results revealed a close relation between the \Hi gas and the assembly history and current structure of stellar disc.
Meanwhile, the bulge seems to be decoupled from the global \Hi properties, but its local relation with the \Hi is still worth investigation with new resolved \Hi data.

\subsection{Relations from \Hi Spectral Stacking}
Recent advances in observational equipments enable the detection of \Hi beyond the redshift of \num{0.1}.
The direct observations of \Hi at $z\sim0.38$ have been shown to be possible with both the interferometer \citep[\qty{178}{\hour} of VLA time,][]{2016ApJ...824L...1F} and single-dish telescope \citep[\qty{95}{\hour} of FAST time,][]{2024ApJ...966L..36X}, but it is still impractical to use them for statistical studies at very high redshift.
Currently, the most successful direct and statistical observation is conducted by the MIGHTEE-\Hi team, who managed to directly map \Hi out to $z\sim0.08$ using MeerKAT \citep{2023MNRAS.525..256P}.
They found no significant change in \Hi mass function \citep{2023MNRAS.522.5308P}, baryonic Tully--Fisher relation \citep{2021MNRAS.508.1195P}, and \Hi size--mass relation \citep{2022MNRAS.512.2697R} over the past billion years, but their statistics are still rather limited with only \num{\sim300} galaxies detected due to the small field.

\begin{figure}
  \centering
  \includegraphics[width=2.2in]{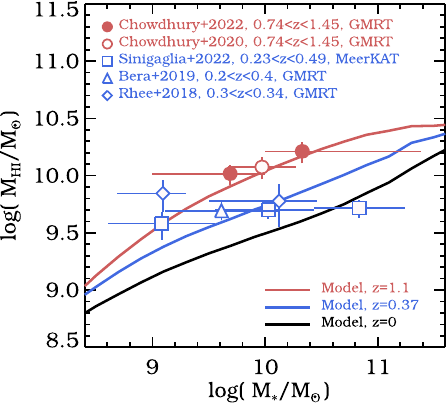}
  \caption{
    A summary of the current \Hi stacking results at different redshifts (symbols), compared with the prediction by the model of \citet{2023ApJ...955...57G}.
    From the Figure~5 of \citet{2023ApJ...955...57G}.
  }
  \label{fig:highzstack}
\end{figure}

Although it is possible to detect \Hi at higher $z$ through gravitational lensing \citep{2016AJ....152...30H}, statistical studies at a higher redshift require the spectral stacking technique, which gives the average \Hi mass among the stacked sample, usually star-forming galaxies.
MIGHTEE-\Hi team conducted stacking studies at $z\sim0.35$ \citep{2022ApJ...935L..13S}, and using the Giant Metrewave Radio Telescope (GMRT), the redshift around \num{1} has been reached \citep{2016ApJ...818L..28K,2018MNRAS.473.1879R,2020Natur.586..369C}.
At $z\sim1$, the \Hi mass of galaxies with $\qty{1e9}{\Msun}\lesssim M_*\lesssim\qty{1e10}{\Msun}$ is systematically larger than the $z=0$ value by \qty{\sim0.55}{\dex}, with a similar slope of \MHi--$M_*$ relation \citep{2022ApJ...941L...6C}.
Meanwhile, the \Hi depletion time is systematically lower by a factor of \numrange{2}{4}, requiring the gas accretion being very active since then \citep{2022ApJ...941L...6C}.
Within this GMRT sample, galaxies at $z\approx1.3$ also have a higher \Hi content, which accounts for \qty{\sim70}{\percent} of total baryon mass, in comparison to the $z\approx0$ value of \qty{33}{\percent} from the xGASS sample \citep{2022ApJ...935L...5C}.
At $z\sim0.35$, however, both MIGHTEE-\Hi team \citep{2022ApJ...935L..13S} and GMRT team \citep{2023ApJ...950L..18B} found a significantly flatter slope of \MHi--$M_*$ relation compared to the $z=0$ one, but with a \qty{\sim0.5}{\dex} discrepancy between each other.
The former crosses the local one from xGASS at $M_*=\qty{1e10}{\Msun}$, with the latter at $M_*=\qty{1e9}{\Msun}$, indicating that the stacked relations still have a large systematic uncertainty.
The \Hi mass function from GMRT data shows the lack of high \MHi galaxies \citep{2022ApJ...940L..10B}.
One possible reason could be the decrease of \Hi due to high star-forming efficiency (SFE), which would then be recuperated through gas accretion.
A summary of these high-$z$ \Hi stacking results is given in Figure~\ref{fig:highzstack}.

Interestingly, the high-$z$ stacking results reveal a downward deviation of the \fHi--sSFR relation from the local one \citep{2022ApJ...941L...6C,2022ApJ...935L..13S}, which implies an evolution in the efficiency of \Hi fuelling the star formation.
It also indicates a possible problem of applying local photometric \Hi-mass estimator to higher redshifts assuming a constant \Hi--colour relation \citep[e.g.][]{2021A&A...648A..25Z}.
By doing so, the \Hi mass may be overestimated, resulting in an overestimated normalization of the photometrically estimated \Hi mass function (HIMF)\@.
The systematic bias of photometric HIMF shape, however, is sensitive to the redshift evolution of \fHi--sSFR slope, which is inconclusive in the literature so far \citep{2022ApJ...941L...6C,2022ApJ...935L..13S}.

Nonetheless, the cosmic variance should not be ignored, especially the $z\sim0.35$ stacking using GMRT, which has only \num{\sim500} galaxies to stack in one single field of $\ang{\sim1}\times\ang{1}$.
When plotting the result from spectral stacking as the cosmic \Hi density, there is still a gap between them and the higher-$z$ absorption results \citep{2020Natur.586..369C}.
A larger sample is required to fully depict and explain the evolution of \Hi properties across a large redshift range.

Stacking could also be used to dig out the faint signal at the low redshift.
By stacking the ALFALFA data cube with the \emph{a~priori} information from galaxy group catalogue, the average \Hi content of the central galaxy or all satellite galaxies could be obtained for a given group halo mass bin.
For satellite galaxies, the total \Hi mass increases with both the group richness and halo mass, while the \MHi of central galaxies show a non-monotonous relation with these two parameters, indicating the dependence on the assembly history, and the role of mechanisms like merging and shock-heating \citep{2020ApJ...894...92G}.
If separating the central galaxies into star-forming and quenched ones, their respective average \MHi's have a very similar relation with either stellar or halo mass, except that the quenched galaxies have a systematically \qty{\sim0.6}{\dex} lower \MHi \citep{2021ApJ...918...53G}.
This indicates the more direct role of \Hi content on the galaxy quenching process than the stellar or halo mass.
Still, it should be noticed that stacking gives the average \Hi mass instead of the median, which is not easily corrected for \citep{2022ApJ...940L..10B}.
Since the distribution of \MHi is not necessarily log-normal, the interpretation and the comparison with direct observations should usually be drawn with caution.

\section{The Kinematics of \Hi}
\label{sec:kinematics}
\subsection{The Flow of \Hi}
\label{ssec:flow}
How \Hi gas flows onto and through the disc is essential to gas accretion and star formation fuelling.
The most direct way to detect and characterize them is to measure the flow velocities.
In the nearby Universe, galaxy star formation and secular evolution tend to be mild and in quasi-equilibrium, and these \Hi flows are expectedly slow in comparison to the bulk circular motion, possibly comparable to the turbulent motion.
Limited by their nature of being weak and the observational limitations, direct measurements of these motions are considered difficult, suffering from strong degeneracies with the geometry of discs \citep{2004ApJ...605..183W}.
A key technical development boosting advances in this field is the 3D fitting with the data cube, instead of 2D fitting with velocity images, which was limited by computational capabilities in the past.

\begin{figure}
  \centering
  \includegraphics[width=3in]{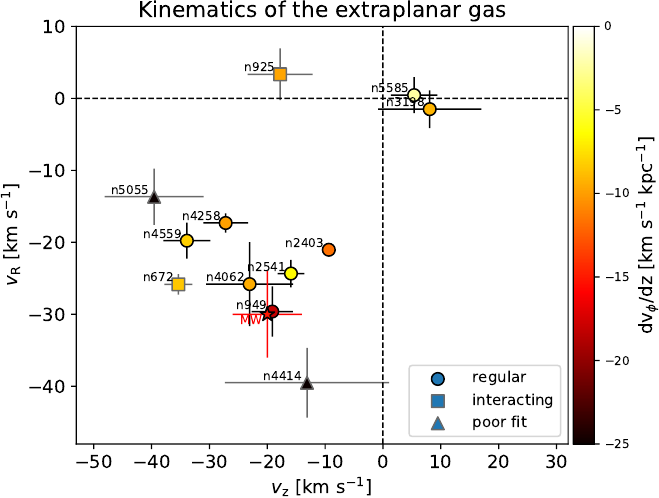}
  \caption{
    Kinematics of the extra-planar \Hi detected in HALOGAS galaxies.
    The values of the radial velocity ($v\tsb{R}$), vertical velocity perpendicular to the disc ($v_z$), and the vertical gradient of the rotational speed ($\mathrm{d}v_\phi/\mathrm{d}z$) are presented.
    Interacting and poorly fitted galaxies are given as squares and triangles.
    Most of the extra-planar \Hi are globally inflowing (located in quadrant~III) with a speed of \qtyrange{20}{30}{\km\per\s}.
    From the Figure~8 of \citet{2019A&A...631A..50M}.
    \copyright~ESO\@.
  }
  \label{fig:extraplanar}
\end{figure}

One possibly important way of gas replenishment in nearby galaxies is the circulation and accompanied additional cooling of gas driven by stellar feedbacks, which is now often referred to as the fountain mechanism.
Over the past two decades, the studies of individual galaxies (e.g.\ NGC~253 \citep{2005A&A...431...65B}, NGC~891 \citep{2005ASPC..331..239F,2007AJ....134.1019O}, NGC~2403 \citep{2002AJ....123.3124F,2023MNRAS.520..147L}, NGC~4559 \citep{2005A&A...439..947B,2017ApJ...839..118V}, and NGC~6946 \citep{2008A&A...490..555B}) consistently support the existence of extra-planar \Hi, its link to the fountain mechanism, and its inflow rate close to the SFR\@.
They typically found that extra-planar \Hi accounts for \qtyrange{10}{20}{\percent} of the total \Hi, has a typical thickness of a few kiloparsec, co-rotates in a lagged way with respect to the thin disc, and flows in the vertical and radial direction with a velocity of \qtyrange{20}{30}{\km\per\s}.
Most of these studies are based on deep \Hi data taken at Westerbork Synthesis Radio Telescope (WSRT) as part of the Hydrogen Accretion in LOcal GAlaxieS (HALOGAS) project \citep{2011A&A...526A.118H}.
Not until recently, such analysis has been conducted on a statistical sample of 15 nearby spiral galaxies with homogeneous data in a coherent way \citep[see Figure~\ref{fig:extraplanar}]{2019A&A...631A..50M}, and obtained similar results and conclusions.
Remarkably, the masses of extra-planar \Hi in this whole sample are in good consistency with the prediction of fountain model, and the net inflow rate after accounting for mass loading is roughly at the SFR level.

The radial flows represent another channel of star-forming gas replenishment, which may start from hot-mode CGM cooling onto the \Hi disc outskirts.
Based on a sample of 54 nearby spiral galaxies and using 3D tilted-ring fitting, \citet{2021ApJ...923..220D} show that \Hi both inflows and outflows radially throughout the disc, with the peak mass flow rates on the level of around \qty{1}{\Msun\per\yr}.
Because flows in opposite directions cancel out, the averaged net inflow rate crossing a wide radial range is typically less than \qty{1}{\Msun\per\yr}, and specifically the average inflow rate outside the optical disc is around one fifth to one tenth of the SFR \citep{2021ApJ...923..220D}.
On the other hand, other studies using different method or data have reached contrary conclusions that the inflow rates are comparable to the SFR \citetext{\citealp{2019ApJ...883...77S,2023A&A...675A..37E}, also see \citealp{2016MNRAS.457.2642S}}.
The speed and direction of radial motions are known to be highly sensitive to systematic uncertainties in the inclination measure, and the azimuthal variation of velocity patterns in response to non-asymmetric disc structures also leads to systematic biases \citep{2004ApJ...605..183W, 2007ApJ...664..204S, 2018A&A...618A.106R}.
The 3D modelling in principle should provide richer information than 2D \citep{2021ApJ...923..220D}, but these factors are not fully accounted for or taken advantage of in any single studies.
Moreover, the role of shocks, which inevitably arise from the crossing of radial and azimuthal gas flows, and the participation of consequent multi-phase gas in the radial inflow remain to be explored.

\begin{figure}
  \centering
  \includegraphics[width=3in]{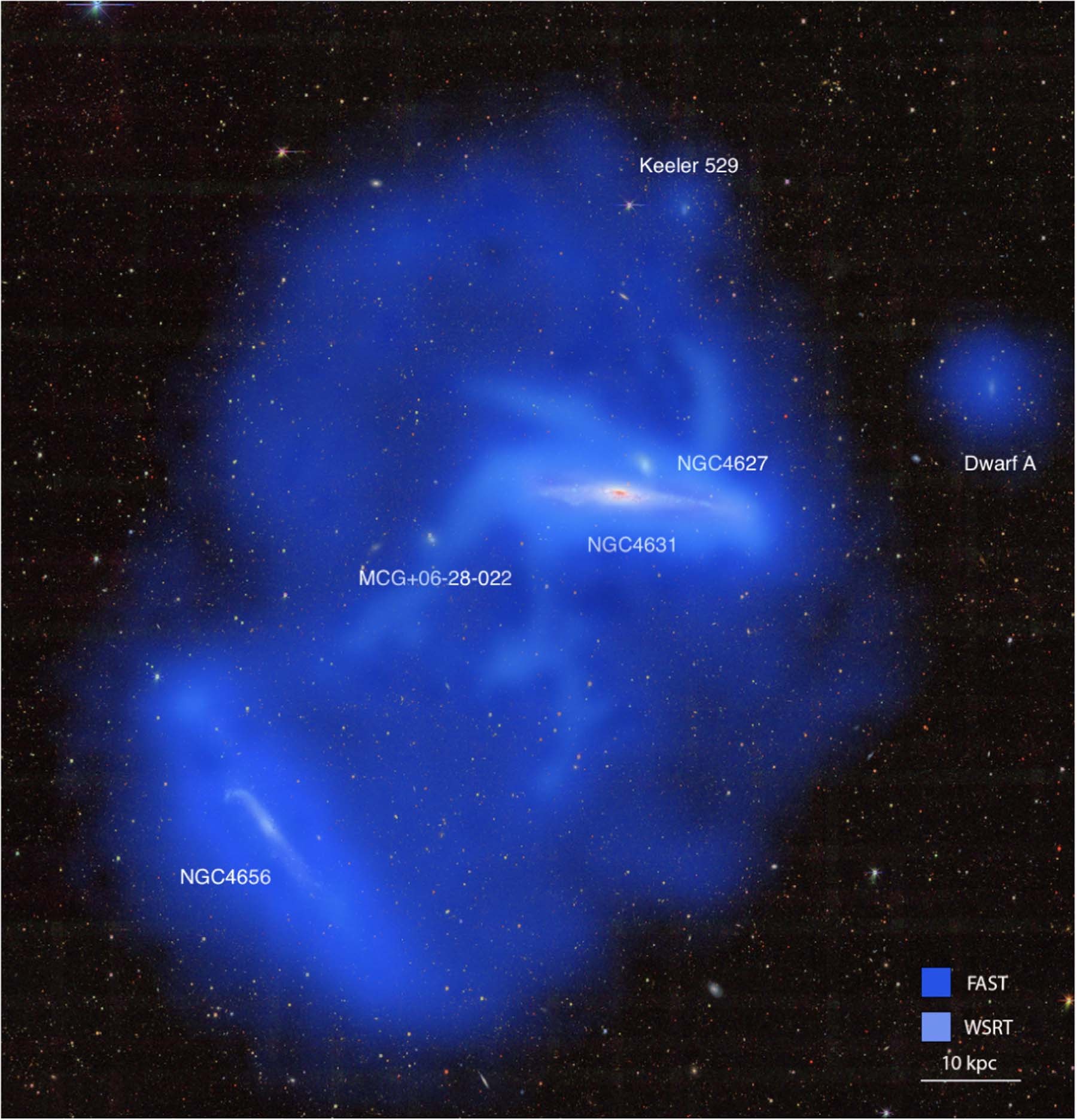}
  \caption{
    A false colour image of the NGC~4631 group and its \Hi gas.
    Light- and dark-blue structures indicate the denser \Hi previously detected by WSRT HALOGAS \citep{2011A&A...526A.118H}, and the diffuse \Hi newly detected by FEASTS using FAST\@.
    The diffuse \Hi accounts for \qty{40}{\percent} of the \Hi in the whole system, and \qty{70}{\percent} beyond the discs.
    From the Figure~1 of \citet{2023ApJ...944..102W}.
  }
  \label{fig:n4631}
\end{figure}

A natural additional origin of \Hi gas is the merger of dwarf galaxies or massive \Hi clouds possibly stripped from dwarf satellites \citep{2019MNRAS.490.4786G}.
With deep data of HALOGAS, it is confirmed that this source of \Hi (down to a mass limit of \qty{1e6}{\Msun}) is unlikely to sustain the SFR \citep{2022A&A...668A.182K}.
However, the studies discussed above in this section are all based on \Hi interferometric images, which have the advantage of high angular resolution, but the disadvantage of missing extended diffuse fluxes.
Recent studies of FAST Extended Atlas of Selected Targets Survey (FEASTS) using moderate-resolution single-dish \Hi images taken by FAST find \qtyrange{10}{50}{\percent} of \Hi missed in the interferometric observation of nearby galaxies, even in HALOGAS \citep{2023ApJ...944..102W,2024ApJ...968...48W}.
One example is shown in Figure~\ref{fig:n4631}.
The missed diffuse \Hi seems to be an analogue or extension of the extra-planar \Hi identified in HALOGAS-based studies within the optical radius, but is likely linked to the strength of tidal interactions in the outer discs.
Because the tidal interaction tends to increase the contacting area and relative velocity between \Hi and the CGM, CGM cooling triggered by thermal conduction \citep{1977ApJ...215..213M} and turbulent mixing \citep{2023MNRAS.520.2148Y} are proposed by these studies as a new potential gas accretion channel that has been overlooked previously.
Determining whether galaxies lose or gain gas through tidal interaction requires a better estimation of the time scales of tidally induced star formation and CGM cooling.
Various \Hi gas fractions in different merging stages have been found observationally \citep{2018MNRAS.478.3447E,2018ApJS..237....2Z,2019ApJ...870..104S,2022ApJ...934..114Y}.
Combining these new observations with new zoom-in simulations may help to understand and constrain the related time scales and contributions from CGM cooling \citep[e.g.][]{2022MNRAS.509.2720S}.

\subsection{The Velocity Dispersion of \Hi}
\label{ssec:disperson}
The velocity dispersion of \Hi reflects the thermal and turbulent motions, which are sustained by dynamic or feedback processes that heat or inject turbulent energy to the gas.
Without a continuous kinetic energy input, the related kinetic energy dissipates within the typical time of crossing \Hi disc thickness, which is typically tens of megayears, close to the lag between star formation and type~II supernova feedback.
Effects from large-scale dynamics on the disc, including gas accretion and radial inflows, are expected to also drive the turbulence in \Hi.
The turbulent dissipation and thermal cooling lead to the condensation of \Hi, the phase transition to molecular gas, and the followed star formation.
Because the velocity dispersions of \Hi vary in a relatively narrow range of \qty{\sim10}{\km\per\s}, a quasi-equilibrium or self-regulation is often expected between turbulence-driven (energy-inputting) processes and star formation \citep[e.g.][]{2018MNRAS.477.2716K}.
The distribution of \Hi velocity dispersion and its correlation with other galactic parameters are thus treated as useful diagnostics for the status of equilibrium and the effects of turbulence-driving mechanisms, particularly stellar feedbacks.

When being azimuthally averaged, the estimation based on SFR surface densities indicates that the supernova feedback should be a sufficient energy source for sustaining the velocity dispersion level of \Hi within the star-forming disc (e.g.\ when the SFR surface density level is greater than \qty{1e-4}{\Msun\per\yr\per\kpc\squared}, \citealp{2009AJ....137.4424T,2020A&A...641A..70B}).
This result holds for both spiral and dwarf galaxies \citep{2013ApJ...773...88S}.
The velocity dispersion of \Hi drops with the increasing radius, and increases with the \Hi surface density, while the ratio between the velocity dispersions of \Hi and CO seems to be rather constant around \num{1.4}, independent of the radius \citep{2016AJ....151...15M}.
However, when one look into the correlation locally, a direct link between the SFR and \Hi velocity-dispersion variance seems to be weak in both spiral and dwarf galaxies \citep{2021AJ....161..175H,2022ApJ...928..143E}.
When the spatially resolved star-formation histories are derived, the \Hi velocity dispersion has the strongest correlation with the SFR \qtyrange{100}{200}{\Myr} ago, instead of the current one \citep{2022AJ....163..132H}.
It seems that, on intermediate scales (azimuthally averaged), there is an injection of turbulent energy from the stellar feedback to the \Hi;
however, most of the feedback energy does not go into local \Hi motion, but is instead captured by other locations or phases followed by quick dissipation or diffusion.
It is also possible that localized quantification is strongly affected by star formation and feedback cycles.

With data of good resolution both spatially and spectrally, the line-of-sight \Hi spectra can be decomposed into narrow and broad components, relating to the cool and warm \Hi \citep{2022ApJ...928..177O}.
The velocity dispersion of both narrow and broad \Hi components drops exponentially with radius, but the former drops faster than the latter, but slower than the molecular gas and SFR surface densities \citep{2015AJ....150...47I}.
The narrow to broad components flux ratio increases more with stellar mass or metallicity, than with localized SFR \citep{2012AJ....144...96I}, and there is a lack of narrow components in dwarf galaxies in comparison to spirals \citep{2015AJ....150...47I}.
They imply the necessity of metal-assisted cooling and shielding for the formation of cool \Hi components.
Interestingly, in dwarf galaxies, the relation of the narrow-component-\Hi surface density with the SFR seems to follow the extension of molecular gas star-formation law at the low-density side \citep{2022AJ....164...82P}, implying a strong link between the cool \Hi and molecular gas.

These results suggest that the velocity dispersion of \Hi in star-forming discs is likely sustained by star formation, but a clear relation may be blurred by the inhomogeneous ISM environment evolving with molecular clouds, star formation, and feedback.
Kinematically cool and warm \Hi components that coexist along the same line of sight potentially carry more information on these interplays.
While the latter may be more closely associated with stellar feedbacks, the former may be the fuelling material for molecular-gas and star formation.

\section{The 3D Structure of \Hi Distributions}
\label{sec:structure}
In the past decade, the radial distribution and thickness of \Hi discs have been measured with diverse samples, derived in different ways, and linked to other galactic properties.

Radially, early studies found the \Hi to follow a tight size--mass relation, where the mass refers to the total \Hi mass, and the size is the characteristic radius for the surface density \SigHi to reach \qty{1}{\Msun\per\pc\squared} \citep[e.g.][]{1997A&A...324..877B}.
Based on a compiled sample of more than 600 galaxies of different types from different surveys, the relation is found to have a small scatter of only \qty{0.06}{\dex} along the size direction, and not to display any shift between galaxies of different optical luminosities, \Hi richness, morphologies, or environments \citep{2016MNRAS.460.2143W}.
The tightness and universality of the \Hi size--mass relation is confirmed with larger and much more homogeneous dataset than before in the past five years \citep{2022MNRAS.512.2697R, 2023PASA...40...32R}.
It is accepted that the size--mass relation does not simply reflect a truncation in \SigHi due to atomic-to-molecular gas conversion at the high-\SigHi side, and ultraviolet (UV) background ionization at the low-\SigHi side, which was considered to be the cause in very early days.
Instead, accurately reproducing it without a systematic shift is not trivial in simulations, and it can be strongly modulated by stellar feedback and other hydrodynamic processes working on the ISM (also see the discussion in Section~\ref{sec:simu}).

The size--mass relation characterizes the distribution of \Hi in the outer disc more than in the inner disc, and the \Hi discs often show flattening or depression at the centre, while bumping or forming a ring at an intermediate radius.
Different models are invoked to characterize the radial distribution in the inner discs, including the atomic-to-molecular conversion-motivated model \citep{2014MNRAS.441.2159W}, the Gaussian-function model \citep{2016A&A...585A..99M}, and the S\'ersic model \citep{2021AJ....161...71H}.
Physically, the flattening or depression of \Hi at a small radius most likely arise from the conversion to molecular gas that directly links with the star formation, which has been observed in spiral galaxies \citep{2008AJ....136.2782L}.
Based on this scenario, it is speculated that the dwarf irregular galaxies observed by LITTLE THINGS \citep[Local Irregulars That Trace Luminosity Extremes, The \Hi Nearby Galaxy Survey,][]{2012AJ....144..134H} project have a substantial amount of dark molecular gas, because these galaxies have similar flattening of \Hi distribution at small radius, while their SFRs are centrally concentrated \citep{2021AJ....161...71H}.

\begin{figure}
  \centering
  \includegraphics[width=3.2in]{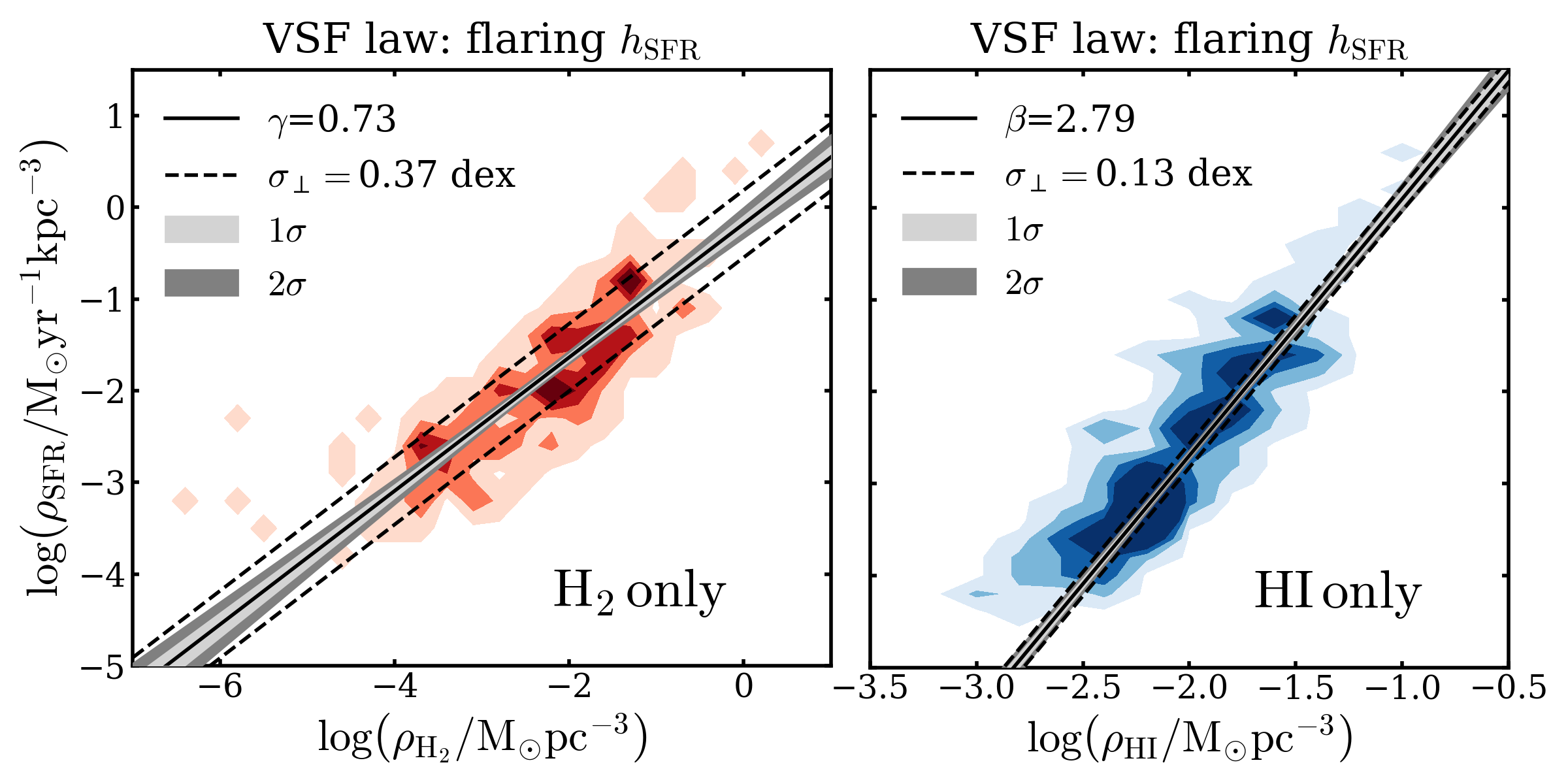}
  \caption{
    The volumetric star-formation law between the SFR and the molecular gas (\emph{left}) or \Hi (\emph{right}).
    The volumetric densities $\rho$ are calculated from the corresponding scale-heights $h$ assuming hydrostatic equilibrium.
    The relation with \Hi is much more tighter than that with H\textsubscript{2}.
    From the Figures 7 and~8 of \citet{2019A&A...622A..64B}.
    \copyright~ESO\@.
  }
  \label{fig:vsf}
\end{figure}

Vertically, following our experience with the MW assuming the hydrostatic equilibrium, the thicknesses or scale-heights of the \Hi are expected to be \qty[input-comparators=\lesssim]{\lesssim1}{\kpc} and to increase with the radius.
They are derived photometrically on edge-on galaxies \citep{2022MNRAS.513.1329Z}, with statistical hydrostatic-equilibrium modelling on less-inclined galaxies \citep{2019A&A...622A..64B, 2020A&A...644A.125B}, or in a hybrid way \citep{2020MNRAS.494.4558Y, 2021ApJ...916...26R}.
The \qty[input-comparators=\lesssim]{\lesssim1}{\kpc} thicknesses and the flaring of \Hi discs have been confirmed in these studies.
The thickness is found to be more strongly correlated with the \Hi mass or \Hi richness, rather than the SFR, sSFR, or SFR surface densities \citep{2021ApJ...916...26R, 2022MNRAS.513.1329Z}.
One possible reason is that the stellar feedback works on the multi-phase gas instead of just \Hi, and the localized potential, cooling, and environmental effects also affect the \Hi thickness \citep{2019A&A...631A..50M}.
The flaring, as well as warping, is found to be the key reason that simply derived \Hi surface densities have weak correlations with the SFR surface densities.
Once these vertical structures have been considered and volume densities have been derived, the \Hi can be even more tightly related to the SFR than the molecular gas in systems assumed to be hydrostatic \citep[see Figure~\ref{fig:vsf}]{2019A&A...622A..64B,2020A&A...644A.125B}.
It is hinted that the volumetric relation may weaken near the galaxy centre, where the hydrostatic assumption may not hold \citep{2020MNRAS.494.4558Y}.

\subsection{The Power Spectrum of \Hi Surface Densities}
The power spectrum (PS) of \Hi surface densities can be a useful tool for the characterization of the \Hi structure in discs, and the identification of major mechanisms driving structure formation \citep{2004ARA&A..42..211E}.
The PS slopes of 2D \Hi emission have been extensively measured for different parts of the MW, as well as other well-resolved galaxies in the Local Group, including the Small Magellanic Cloud (SMC), Large Magellanic Cloud (LMC), and M31 \citep[e.g.][]{2001ApJ...561..264D, 2001ApJ...548..749E}.
We summarize from these early studies that the slopes are typically between \numlist{-2.8;-3.5}, systematically steeper than that of the 2D PS in a turbulent incompressible medium passively transported by the \citet{1941DoSSR..30..301K} velocity field.
Because of the difference and possible (de-)coupling between the density and intensity field, the PS slope of surface density observed in \qty{21}{\cm} emission line is modulated by the velocity fluctuations, and thus differs between thin and thick gas layers \citep{2000ApJ...537..720L}.
The slope values, as well as the characteristic length-scales where the slope changes, are expected to carry information on the complex physics driving the density and velocity fluctuations, including the gas-clouds assembly, feedback injection, shock compression, magnetohydrodynamics modulation, dynamic effects of bars and spiral arms, and the disc warps and flaring.
With new data of better qualities, the power spectra of \Hi have been revisited with new details, both in the MW and other galaxies.

\begin{figure}
  \centering
  \includegraphics[width=2.75in]{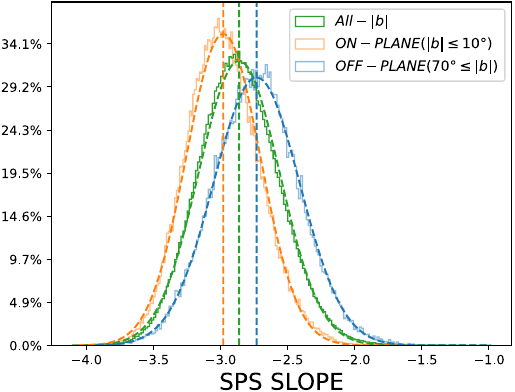}
  \caption{
    The histogram of the spatial power spectrum slope measured for Milky Way \Hi.
    The slope ranges between \numlist{-2.6;-3.2}, and varies with the Galactic latitude.
    From the Figure~6 of \citet{2023ApJ...958..192M}.
  }
  \label{fig:pspec}
\end{figure}

Using the \qty{\sim13000}{\deg\squared} Galactic Arecibo $L$-band Feed Array \Hi \citep[GALFA-\Hi,][]{2011ApJS..194...20P} survey data, it has been confirmed for MW that the slopes of 2D intensity PS are between \numlist{-2.6;-3.2} for the length scales between \ang{;16;} and \ang{20;;}, with the slopes varying mostly with Galactic latitudes and being almost constant in each latitude layer \citep[see Figure~\ref{fig:pspec}]{2023ApJ...958..192M}.
In the LMC, the slopes steepen at small length scales of a few hundreds of parsecs, and additionally steepen close to the \ion{H}{2} regions, implying effects of stellar feedbacks in such low-mass galaxies, which, however, are not observed in the SMC \citep{2019ApJ...887..111S}.
Based on a sample of 33 spiral galaxies from THINGS \citep[The \Hi Nearby Galaxy Survey,][]{2008AJ....136.2563W}, it was found that the \Hi-variance spectrum (similar to PS, but calculated with different scale kernels) has three distinct parts:
an inner part with a bump on scales of a few hundreds of parsecs, an intermediate flat part ending at roughly half the optical radius, and a remaining steep outer part \citep{2021A&A...655A.101D}.
Analysis correlating the characteristic length scales and amplitudes of these features revealed that, these three parts are possibly associated respectively with the strength of star formation and feedback, the atomic- to molecular-dominated disc transition, and large-scale disc dynamics or external environmental perturbations \citep{2021A&A...655A.101D}.
Because of the piece-wise feature of PS, when comparing PS slopes between different studies based on samples at different distances, the resolution and length-scales being considered need be carefully controlled for.
For example, studies focusing on the global (and hence more biased towards large-scale) PS of \Hi in galaxies typically find a lack of correlation between the PS slopes/length scales and stellar feedback, and argue for large-scale gravitational energy input source, such as disc instabilities and self gravity, as the main driver of the turbulence \citep{2020MNRAS.496.1803N, 2023MNRAS.526.4690N}.

To summarize, the new developments in the field of 2D \Hi PS measurements include the extension in sample diversity and a better combination with multi-wavelength data, helping to directly identify and/or characterize possible driving physical mechanisms.
This field seems to be reviving after a period of relative silence.
We look forward to more comprehensive studies on the correlation of PS features with parameterized physical factors, as well as on the interplay between different length-scales.

\section{The Role of \Hi Gas in Environmental Studies}
\label{sec:env}
It has been well established since \citet{1980ApJ...236..351D} that galaxy properties (including morphology, colour, star formation activities, etc.)\ depend on the group environment, which is correlated with the local galaxy density to the first order.
The evolution of galaxies could be altered by environments through different mechanisms, which has been largely summarized as the hydrodynamical ones and gravitational ones \citep{2022A&ARv..30....3B}.
The \Hi structure is usually much more extended than stellar disc \citep{2016MNRAS.460.2143W} and thus susceptible to deformation by environmental processes, especially at the outskirts where the surface density is low.
As a result, \Hi could be utilized as a sensitive tracer to ongoing or past environmental interactions.
As it also serves as the reservoir of fuel for star formation, \Hi gas could help to determine the timescale of the environment reorienting the galaxy evolution.

\begin{figure}
  \centering
  \includegraphics[width=3in]{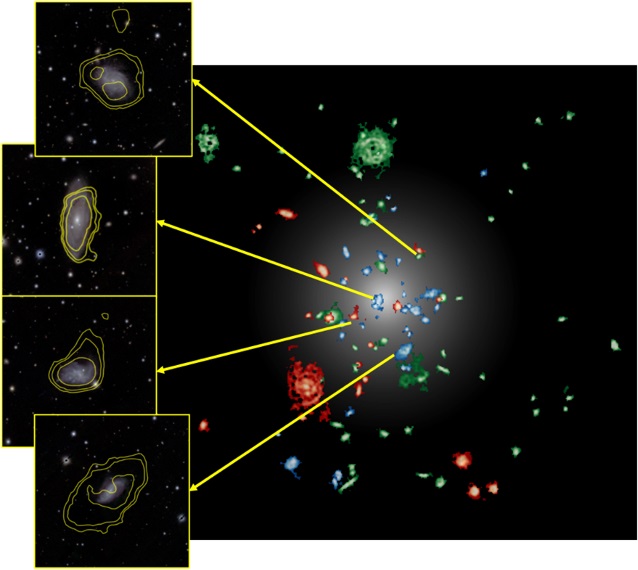}
  \caption{
    WALLABY-detected galaxies in the Hydra cluster.
    The disc sizes are not to the scale.
    Four galaxies possibly undergoing RPS are shown as optical images overlaid by \Hi contours.
    Modified from the Figure~1 of \citet{2021ApJ...915...70W}.
    \copyright~AAS\@.
  }
  \label{fig:hydramap}
\end{figure}

With the aforementioned xGASS sample spectra, it is found that satellite galaxies exhibit more asymmetric \Hi spectra than centrals \citep{2020MNRAS.492.3672W}.
In addition, past several years have seen the start of several new \Hi observational campaigns, providing data with higher resolution and sensitivity covering a sky area larger than ever.
Several new \Hi sources are detected individually showing features of tidal interactions \citep[e.g.][]{2019MNRAS.488.5352K,2023MNRAS.521.5177N} or ram-pressure stripping \citep[RPS, e.g.][]{2019MNRAS.487.2797E,2023A&A...676A..92B}.
Moreover, it is now possible to achieve the blind mapping of \Hi in a whole group or cluster with moderate resolution and sensitivity.
The observations are not limited to nearby massive clusters (like the Virgo) targeting pre-selected sources \citep[e.g.\ VIVA, VLA Imaging of Virgo in Atomic gas,][]{2009AJ....138.1741C}.
Instead, they have contiguously covered systems like Hydra~I cluster \citep[see Figure~\ref{fig:hydramap}]{2021ApJ...915...70W}, Abell~2626 \citep{2021A&A...654A.173H}, Fornax cluster \citep{2023A&A...673A.146S}, Eridanus supergroup \citep{2021MNRAS.507.2300F}, etc., even out to several virial radius.
With the high sensitivity of FAST, the possibility of detecting tidally induced intragroup \Hi in both nearby \citep[such as the M106--NGC~4288 pair at \qty{\sim8}{\Mpc},][]{2021ApJ...922L..21Z} and more distant systems \citep[such as the Stephan's Quintet at \qty{\sim85}{\Mpc} away,][]{2022Natur.610..461X,2023ApJ...954...74C} has also been highlighted, but the limited resolution restricted a detailed analysis and concrete interpretation in the latter case (where the resolution is \qty{\sim100}{\kpc}).

These various new datasets provide the possibility to statistically construct the evolution of galaxies infalling into a dense environment.
In the massive and hot Hydra~I cluster, ram pressure is found to be highly prevalent (affecting \qty{70}{\percent} of the infalling \Hi-rich population within \num{1.25} times the virial radius) and very effective (\qty[input-comparators={\lesssim}]{\lesssim200}{\Myr}) in removing `strippable' \Hi gas, but it need a much longer time to have considerable cumulative effects \citep{2021ApJ...915...70W,2022MNRAS.510.1716R}.
In the similar-mass Abell~1367 cluster, \qty{26}{\percent} of late-type galaxies show disturbance in \Hi morphology and kinematics \citep{2018MNRAS.475.4648S}, compared with the fraction of \qty{2}{\percent} among isolated samples.
The rather smaller Fornax cluster has at least six galaxies with evident and long \Hi tail \citep{2023A&A...673A.146S}, indicating the intense environmental influence.
With a slightly lower halo mass, the Eridanus supergroup is found to have a systematically \Hi-deficient satellite population \citep{2021MNRAS.507.2949M}, which is strongly influenced by satellite--satellite tidal interaction and shows corresponding inside-out (instead of outside-in) reddening \citep{2022ApJ...927...66W}.
These studies provided strong evidence for the dominant role of environment in satellite evolution, but also presented the new challenge of quantifying the factors involved.
The cluster halo mass is certainly the first-order determinant.
Cluster substructures have been found to hold galaxies of higher level of \Hi morphological asymmetry \citep{2023A&A...676A.118D}, possibly due to stronger tidal interaction, but the measurements on marginally resolved systems should be interpreted with caution.
Meanwhile, null results have been found for the \Hi spectra of mergers \citep{2022ApJ...929...15Z}, and some simulation results indicate that the ram pressure plays a more important role in producing spectral asymmetry \citep{2020MNRAS.499.5205W}.
We shall keep in mind that the spectral asymmetry contains less information than the morphological one.
Substructures possibly shielding the members from harsh environment \citep{2021A&A...648A..32K} or with a higher \Hi content \citep{2023ApJ...956..148L} have also been found.
Other possible factors include the anisotropy of intra-cluster medium and the virialization state, which need a larger cluster/group sample to quantify.

The fate of a satellite galaxy further depends on the initial condition when it fell into its current cluster, which is in turn shaped by the previous evolution in a smaller structure, or the so-called `pre-processing'.
New \Hi observations made it possible to evaluate the role of these less-massive galaxy groups, many of which are infalling into a larger cluster.
Fornax~A (NGC~1316) group, the mass of which is ${\sim}1/3$ of the Fornax cluster, has already hold galaxies showing significant \Hi disc truncation \citep{2021A&A...648A..32K}.
In NGC~4636 group, some galaxies at ${\sim}2\Rzoo$ are found to already be \Hi-poor \citep{2023ApJ...956..148L}.
Therefore, it is possibly more instructive to understand `pre-processing' in a relative sense, and view the galaxy evolution in the context of continuous assembly of cosmological structures.

One interesting difference between the lower-mass group and the more-massive clusters is that the hydrodynamical processes are on average weaker in the former \citep{2021PASA...38...35C}, making it more likely to have different environmental effects at work simultaneously.
Weaker environment is less likely to induce significant morphological change in \Hi structure.
Therefore, identifying the major mechanism for each galaxy is difficult.
One solution is to infer the strength of stripping mechanisms from the satellite coordinates and redshifts in respect of the cluster center, which has been used for RPS \citep{2015MNRAS.448.1715J,2017ApJ...838...81Y,2020ApJ...903..103W,2021ApJ...915...70W}.
\citet{2023ApJ...956..148L} further introduced a method to determine the fraction of gas being strippable by tidal interaction, utilizing the relative position and velocity from neighbour satellites, and succeeded in disentangling the RPS and satellite tidal interactions in NGC~4636 group.
They found that \qty{41}{\percent} of \Hi-detected satellites in the group are simultaneously influenced by both processes, and the cooperation between them possibly make the stripping more efficient.
It still needs more work to fully capture the complex coexisting environmental effects and break the degeneracy.

One major obstacle to observationally analysing the environmental processing is the degeneracy and uncertainty from the projection, which is important for inferring the galaxy orbits and infalling histories.
The only orbital observables we could obtain are the celestial position and line-of-sight velocity.
Mock observations suggested that the projective and 3D values of the velocity and distance relative to group/cluster centre could have a median offset \qty{>0.1}{\dex} with a \qty{>0.25}{\dex} scatter \citep{2023ApJ...956..148L}, largely influence the calculation of ram pressure and other processes.
For individual galaxies, RPS-driven \Hi morphology could serve as the indicator for orbital orientation \citep{2023A&A...673A.146S}, but sometimes, it is still difficult to differentiate several possible scenarios \citep{2022A&A...668A.184H}.
Statistically, the projected phase-space diagram is useful for estimating the infalling stage of member galaxies \citep{2021MNRAS.500.1784D}, and \Hi richness could be used to select out probable first-time infallers \citep{2023ApJ...956..148L}, which is helpful for deducing the initial condition.

Due to the aforementioned degeneracies and uncertainties, it is necessary to understand in a statistical way the on-going galaxy infalling, stripping, and quenching processes.
An illuminating methodology is to use simplified analytical model to fit the general trend of satellite evolution using \Hi excess/deficiency and/or projected phase-space position as the clock.
\citet{2006ApJ...651..811B} proposed analytical models of starvation and RPS, and compared the prediction to the observed distribution of \Hi deficiency and SFR suppression at a given stellar mass in the Virgo cluster \citep{2014A&A...570A..69B,2023A&A...669A..73B}.
They concluded that starvation alone cannot produce the red and low-SFR satellite population.
\citet{2023ApJ...956..148L} estimated the fraction of \Hi that is susceptible to stripping processes for each satellite in a \qty{\sim1e13}{\Msun} group.
Using the projective group-centric distance as a proxy of infalling time, they estimated the typical stripping timescale of strippable \Hi, which becomes as short as the \Hi cloud free-fall time (\qty{\sim50}{\Myr}) within $0.5\Rzoo$, and the \Hi-depletion timescale, which strongly depends on the initial \Hi richness crossing the group virial radius.
Alternatively, \citet{2023A&A...675A.108K} describe the \Hi removal process in Fornax cluster using the satellite accretion rate of the cluster and the halving time of \Hi.
Comparing the modelled and observed \MHi--$M_*$ scaling relation, they reported an \Hi halving time of \qty{\sim30}{\Myr}.
With their own assumptions, each of these models focuses on different aspects, and need future expansion to more systems.

\section{\Hi in Hydrodynamic Simulations}
\label{sec:simu}
As a major ISM component, the observed \Hi properties and statistical metrics have been compared with the predictions from hydrodynamic simulations.
However, such a comparison is not easy.
As pointed out in many recent insightful reviews on galaxy models and simulations, gas properties are difficult to reproduce because it is necessary to rely on many sub-grid physical models so as to lower the computational costs \citep{2015ARA&A..53...51S,2017ARA&A..55...59N, 2023ARA&A..61..473C}.
Additionally, except for a few zoom-in suites, \Hi is predicted from the \qty{1e4}{\K} total gas in a post-processing way in most hydrodynamic simulations with a cosmological background.
We remind the reader again that so far, observations of detailed \Hi distribution and kinematics are limited to biased samples, often the \Hi-rich ones.
Despite the limitations, comparing the observed \Hi with simulated predictions has proved useful in identifying the influential physical processes and constraining the whole galaxy evolution model.

The \Hi disc size at a given \Hi mass is systematically over-estimated in the TNG100 and TNG300 \citep{2019MNRAS.487.1529D}, consistent with the over-predicted central deficit in \Hi in these simulations as well as in TNG50 \citep{2023MNRAS.521.5645G, 2023ApJ...957L..19S}.
These analyses consistently point to the thermal-mode AGN feedback in TNG continuously injecting energy into the low-mass \Hi-rich galaxies that is too efficiently at ionizing the gas near the galaxy center.
Concerns about the AGN feedback model of TNG simulations are also raised from the \Hi scaling relations.
In TNG100, the star-formation-quenched galaxies retain too much \Hi gas for their stellar masses, because the AGN feedback mostly redistributes the cold gas to large radius, instead of depleting the global cold-gas reservoirs \citep{2022ApJ...941..205M};
the satellites in halos more massive than \qty{1e12}{\Msun} have too little \Hi at a given stellar mass, because they are strongly affected by the kinetic-mode AGN feedback of central galaxies injecting energy in a pulse, directed fashion, but are unable to re-accrete the ejected gas \citep{2019MNRAS.483.5334S, 2020ApJ...894...92G}.

In simulations requiring a post-processing for separating the \Hi from the total neutral gas, the recipes can be roughly divided into two categories, dependent on the mid-plane pressure \citep{2008AJ....136.2782L} or on the dust and radiation field \citep{2011ApJ...728...88G}, respectively.
These two recipes seem to predict similar surface density distributions of \Hi in the Auriga \citep{2017MNRAS.466.3859M}, unlike the large offsets between the two recipes predicted in the earlier Evolution and Assembly of GaLaxies and their Environments (EAGLE) that has a lower resolution and less matured stellar feedback models \citep{2016MNRAS.456.1115B}.
On the other hand, in Auriga, the scale heights at a given SFR are different between the two recipes, with the mid-plane-pressure--dependent partition predicting thinner \Hi discs \citep{2017MNRAS.466.3859M}.

Reproducing the observed thin disc of \Hi seems to prefer star-formation models with a gas turbulence and virial-status-dependent SFE, and stellar feedback models with early-stellar feedbacks, such as stellar winds, photoionization, photoelectric heating, and radiation pressure from OB stars \citep{2024MNRAS.531.1158G}, which are key features of the FIREbox, part of the Feedback in Realistic Environments (FIRE) project \citep{2014MNRAS.445..581H}.
On the other hand, the Auriga and Illustris seem to significantly over-predict \Hi-disc thicknesses \citep{2017MNRAS.466.3859M,2019MNRAS.486.4686K}, possibly related to the significant vertical gas accretion through the fountain mechanism, after the modelled AGN feedback heats gas to large radius through injecting hot bubbles.
Auriga also predicts a positive correlation between SFR and the \Hi scale height \citep{2017MNRAS.466.3859M}, which is not supported by observations \citep{2021ApJ...916...26R, 2022MNRAS.513.1329Z}.
Despite the success in predicting the \Hi thin discs as well as their flaring feature, FIREbox tends to systematically over-predict the \Hi disc size at a given \Hi mass, and over-predict the \Hi surface density throughout the inner discs \citep{2024MNRAS.531.1158G}.
The causes are not identified so far, raising a warning about complex interplay between the baryonic physical processes.

There is a great deal of attention on reproducing the \Hi-data--based Tully--Fisher relation.
The rotational velocities of \Hi are under-predicted for intermediate-mass galaxies in earlier zoom-in simulations of the FIRE project \citep{2018MNRAS.473.1930E} and in the Simba \citep{2020MNRAS.498.3687G}, and over-predicted in low-mass galaxies in A Project On Simulating The Local Environment (APOSTLE) and EAGLE \citep{2017MNRAS.464.2419S}.
Simba further predicts a flattening of baryonic mass at the high-velocity end \citep{2020MNRAS.498.3687G}, which does not match the latest observation of massive spiral galaxies \citep{2021MNRAS.507.5820D}.
Reasons for the discrepancies in simulated and observed Tully--Fisher relations seem to relate to the processes regulating the galaxy-formation efficiency in dark matter halos and baryonic mass--size relation \citep{2017MNRAS.464.4736F}, which are long-lasting problems of simulations, reflecting more-general uncertainty and complex degeneracy of implemented baryonic physical models \citep{2023ARA&A..61..473C} rather than those specifically with \Hi distribution and kinematics.

\section{Summary and Near-future Perspectives}
\label{sec:summary}
Hereby, we reviewed the major advances and new results of the \Hi observations and studies in the past decade.
The major points are summarized as follows:
\begin{enumerate}
  \item The statistical \Hi scaling relations have been pushed to a stellar mass of \qty{1e9}{\Msun} with xGASS\@.
        The \Hi within the stellar disc is confirmed and characterized as an intermediate stage of star formation cycle.
        Spectral stacking has reached the redshift of \num{0.1} and found signatures of the scaling-relation evolution.
  \item With the advance of 3D fitting technique, the \Hi radial flow within the disc could be estimated for nearby galaxies, but could not fully account for the SFR\@.
        The extra-planar \Hi was analysed statistically, and was found to be consistent with the fountain model.
        The newly detected large-scale diffuse \Hi may also be a channel of gas accretion.
        The \Hi velocity dispersion is likely from the energy released by star formation.
        A full energy budget and more details about the distribution among different gas phases, including the warm and cool phases of the \Hi, may provide more insight in the future.
  \item The \Hi size--mass relation holds in a larger dataset.
        The depression of the density around disc centre is possibly related to the conversion to molecular gas, but additional effects, including the stellar feedback and disc dynamics, require further investigation.
        The radial profile of \Hi thickness is measured under hydrostatic assumption, and could reflect the relation between gas and star formation.
        The power spectrum is a promising tool for the analysing of physical mechanisms within ISM\@.
        These analyses are biased to very \Hi-rich spiral and dwarf irregular galaxies, and need be extended to more-general samples.
  \item \Hi has been a successful environmental tracer.
        New surveys could provide the blind mapping of a whole high-galaxy-density structure for the statistical analysis of satellite evolution.
        Currently, the major obstacles are the degeneracy of coexisting environmental processes, progenitor and orbital uncertainties, and the observational projection effects.
  \item Contemporary simulations are still struggling to reproduce the details of \Hi properties, but they could be very useful in testing the possible factors at work.
\end{enumerate}

With the rapid advancements in large surveys, we will be in a much more advantageous position to revisit many of the topics discussed in this review.
We will focus on surveys at the FAST telescope to make examples here.
Insightful future perspective for the SKA and SKA pathfinder telescopes can be found in the many reviews listed in the Introduction of this paper.

With deeper surveys like the FAST All Sky \Hi (FASHI) survey \citep{2024SCPMA..6719511Z} and the Commensal Radio Astronomy FAST Survey \citep[CRAFTS,][]{2019SCPMA..6259506Z} starting to release data, \Hi scaling relations can be extended to galaxies of lower masses or poorer \Hi richness.
Both surveys will cover twice larger sky than the previous ALFLAFA, but they have distinct scientific advantages.
CRAFTS has a relatively contiguous sky coverage and uniform depth, while FASHI goes deeper in a few regions.
CRAFTS will thus be suitable to support advances in statistical studies, extending the \Hi mass function and scaling relations to lower \Hi masses, making them less suffering from cosmic variance at low redshift (see discussion of such limitation of ALFALFA in \citealp{2023ApJ...955...57G}), and enabling a larger number of galaxy property bins for study of secondary dependence.
Studies with FAST-observed \Hi data in this direction, though not utilizing the CRAFTS dataset yet, have been conducted, constraining evolution of red galaxies \citep{2022MNRAS.516.2337W}, bulge-dominated galaxies \citep{2024ApJ...963...86L}, merging luminous infrared galaxies (LIRGs) \citep{2022ApJ...929...15Z}, major-merger galaxies \citep{2022ApJ...934..114Y}, and the relation between the atomic-to-molecular ratio and properties of the ionized gas \citep{2024arXiv240319447Y}.
It is worth pointing out that, in addition to the integral fluxes indicating \Hi mass and the line width indicating dynamic mass, the spectral shape carries information on the radial and non-axisymmetric distribution of the \Hi in the galaxy, for which parametrization tools have been developed, but science values remain little exploited \citep[e.g.][]{2022ApJS..261...21Y,2022ApJ...930...85Y,2023ApJ...950..163P}.
The stacking technique has proved a tool that is never out of date, as it always provides a preview of the fainter side, where the existing survey depth gets close to but does not directly reach \citep[e.g.][]{2020ApJ...894...92G,2021ApJ...918...53G,2022ApJ...933L..12G}.
It is most easily applied to homogeneous surveys with relatively uniform noise levels like CRAFTS\@.
We look forward to new clues on star formation quenching, gas depletion in groups and clusters, AGN and stellar feedback effects, and formation of extremely low-mass and low-surface brightness galaxies, which are topics highly relying on the relatively faint side of \Hi observations.

On the other hand, FASHI will be useful for searching for rare sources.
FAST has demonstrated this capability in the past years.
It found the candidate dark galaxy, Cloud~9 around the galaxy M94 \citep{2023ApJ...952..130Z}, the large amount of \Hi and long \Hi tail around the compact group, the Stephan's Quintet \citep{2022Natur.610..461X}, and many other new tidal features \citep[e.g.][]{2021ApJ...922L..21Z}.
The existence and fate of these \Hi clouds and structures in hot gas far away from galaxies arouse many interests in follow-up observational and theoretical studies \citep[e.g.][]{2023ApJ...956....1B}.
For many years, the \Hi bridges of the M81 system have been used to demonstrate the \qty{1e20}{\per\cm\squared} \Hi imaging power in revealing the optically dark link between galaxies \citep{1994Natur.372..530Y}.
Now with FAST and FASHI, such links are found prevalent in \Hi-rich moderate- to high-density regions when observed at \qty{\sim1e18}{\per\cm\squared} column density level.
A galaxy zoo built for the FASHI dataset will be potentially useful for boosting discoveries.

The targeted \Hi imaging capability of FAST will remain powerful for the coming decades, because it has the highest resolution as a single-dish telescope, and a relatively low level of side lobes \citep{2020RAA....20...64J}.
This total power imaging capability is valuable for Local Volume systems that tend to have large-scale \Hi structures.
FEASTS is a promising endeavour in this direction, as it provides a relatively large field of view and a uniform mapping of northern-sky galaxies selected mostly by the \Hi flux \citep{2023ApJ...944..102W,2024ApJ...968...48W}.
By directly mapping the \Hi of galaxies down to column density levels \qty[input-comparators=\lesssim]{\lesssim1e18}{\per\cm\squared}, we may reveal the ISM--CGM interface, bridge the gap between the ISM and the low-column density CGM/IGM regime, which is detected by optical and ultraviolet absorption lines.
These new observations will provide key insight on galaxies interacting with the CGM, companion galaxies, group environment, and larger-scale structures, particularly the gas accretion and feedback processes as engines of baryonic flows.

\section*{Acknowledgements}
The authors thank the anonymous referees for their constructive and helpful comments.
JW thanks the support by National Key Research and Development Program of China (Grant No.~2022YFA1602902), National Science Foundation of China (Grant Nos.~12073002, 12233001, and 8200906879), and the China Manned Space Program.

\bibliography{ms}
\clearpage
\end{document}